\documentclass{XrU2005}

\usepackage[]{graphicx}
\usepackage[]{epsfig}

\title{Extended sources in the XMM-Newton slew survey}
\author[1]{V. Lazaro}
\author[1]{R. Saxton}
\author[2]{A.M. Read}
\author[1]{M.P. Esquej}
\author[3]{M.J. Freyberg}
\author[1]{B.Altieri}
\author[1]{D. Bermejo}
\affil[1]{European Space Agency (ESA), European Space Astronomy Centre (ESAC), Villafranca, Apartado 50727,  28080 Madrid, Spain}
\affil[2]{Dept. of Physics and Astronomy, Leicester University, Leicester LE1 7RH, U.K.}
\affil[3]{Max-Planck-Institut f\"{u}r extraterrestrische Physik, Giessenbachstrasse 1, 85784 Garching, Germany}

\begin{document}


\maketitle

\begin{abstract}
The low background, good spatial resolution and great sensitivity of the EPIC-pn camera on XMM-Newton give useful limits for the detection of extended sources even during the short exposures made during slewing maneouvers. In this paper we attempt to illustrate the potential of the XMM-Newton slew survey as a tool for analysing flux-limited samples of clusters of galaxies and other sources of spatially extended X-ray emission.  
\end{abstract}

\section{Introduction}

The XMM-Newton slew survey project has currently catalogued of the order of 4000 sources from 15\% of the sky, with a limiting flux of $\sim10^{-12}$ ergs s$^{-1}$ cm$^{-2}$ in the 0.2--12 keV energy band \citep{freyberg}. Up to 20\% of these sources are reported to be extended by the source detection software. While much work remains to be done on removing spurious sources, many identifications can already be made with known clusters of galaxies, groups of galaxies, nearby galaxies or supernova remnants.

At the flux levels probed here a significant fraction of sources are expected
to be extended.
In the Einstein slew survey \citep{elvis}, which covered $\sim50$\% of the sky,
 143 extended sources (clusters, 
galaxies and SNR) were found representing 17\% of the detections. In the deeper ROSAT bright source catalogue \citep{Voges} $\sim6$\% of the uniquely identified sources were found to be galaxy clusters.

\section{Processing}
A description of the general processing steps for slew data together with the 
solution of the particular attitude reconstruction issues has been presented
elsewhere \citep{Read,Saxton}. The source detection pipeline, consisting of a chain of
the {\it EMASK}, {\it EBOXDETECT}, {\it ESPLINEMAP} and {\it EMLDETECT} tasks,
has been tuned to detect point sources in the very low background conditions usually associated with slew exposures. An assessment has been made on slew images containing known extended sources, to find the best parameters to use within this pipeline to detect extended features. This showed that the 
parameters currently used are also optimal for detecting extended sources
up to a diameter of a few arcminutes, with the caveat that a 'Beta' model 
gives a better fit to the spatial profile of clusters of galaxies than the
default 'Gaussian' model.

\section{Results}

\begin{figure}[ht]
\centering
\rotatebox{-90}{\includegraphics[width=0.8\linewidth]{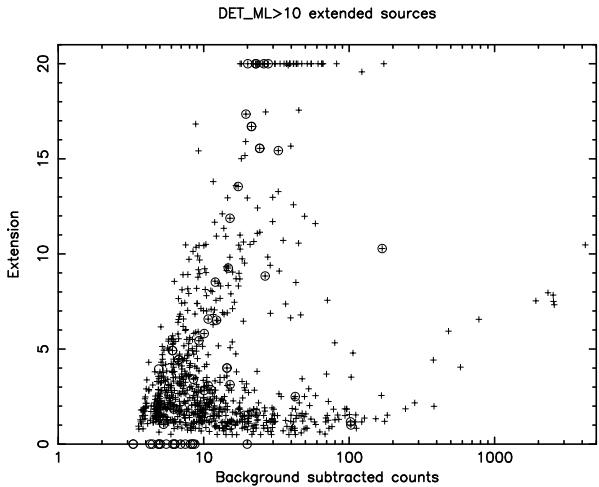}}
\caption{A plot of background subtracted counts against the extension parameter, measured in units of 4.1 arcsecond pixels and limited to 20 pixels. Points coincident with known clusters are circled, although some clusters apparently detected with zero extension, may in fact be coincidences with unrelated point sources. The branch showing increasing extension for very bright sources  is due to pile-up effects distorting the radial profile. \label{fig:extcnts}}
\end{figure}

In Figure~\ref{fig:extcnts} the high significance (DET\_ML $>10$) extended
sources are shown with the extension parameter plotted against the number
of detected background subtracted counts. Sources associated with known Abell and Zwicky catalogue clusters are circled and many detections remain to be identified.

\subsection{Clusters of Galaxies}

\begin{figure}[ht]
\centering
\includegraphics[width=1.0\linewidth]{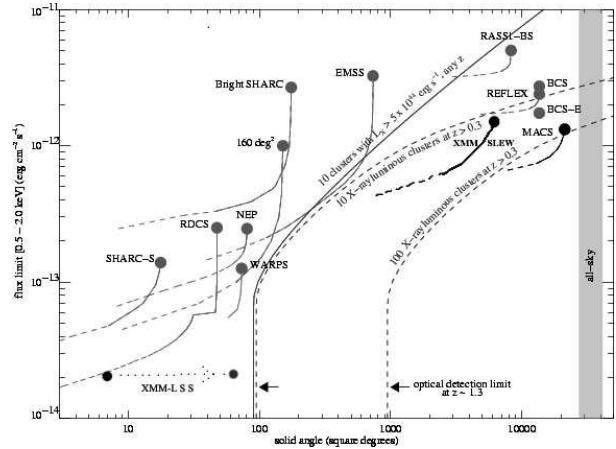}
\caption{A plot of cluster survey depths v area. The XMM slew curve will move to the
right as more slews are accumulated. Other surveys are based on ROSAT data with the exception of the EMSS and XMM-LSS (see \citet{pierre}). \label{fig:depth}}
\end{figure}

The slew survey covers a large sky area to a depth which is comparable with some of the better previous all-sky X-ray cluster surveys (Fig.~\ref{fig:depth}). 
In the regions of overlapping slews, such as the ecliptic poles, co-adding data will lead to a deeper survey albeit over a smaller area.

\begin{figure}
\centering
\includegraphics[width=0.8\linewidth]{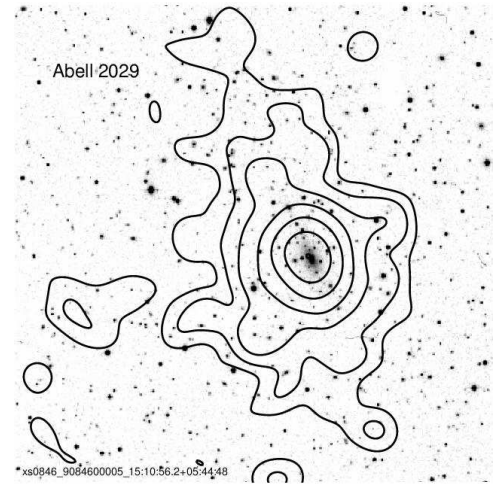}
\caption{X-ray contours of a smoothed slew image of Abell 2029 overlaid on a digital sky survey image. 
\label{fig:cluster}}
\end{figure}

Cross-correlations with the Abell and Zwicky catalogues show 55 coincidences of slew sources with known clusters. Of these, 37 are detected as extended objects. The very brightest examples contain more than 100 counts, sufficient to show some cluster morphology (Fig.~\ref{fig:cluster}). 

%\begin{figure}
%\centering
%\includegraphics[width=0.8\linewidth]{abell3581.ps}
%\caption{Comparison of cluster survey areas and depths. \label{fig:depth}}
%\end{figure}

\subsection{Supernova Remnants}
\begin{figure}
\centering
\begin{tabular}{c}
{\includegraphics[height=6cm]{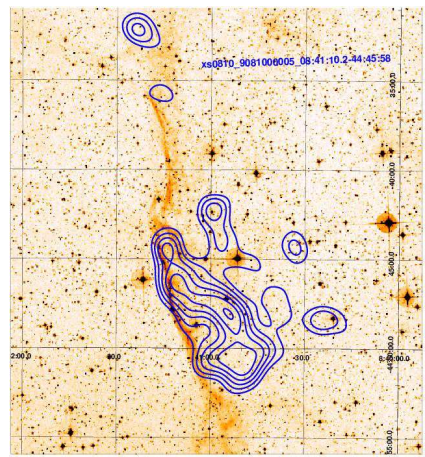}}
\end{tabular}
\caption{X-ray contours of the Vela supernova remnant
overlaid on a DSS image. \label{fig:snr}}
\end{figure}

A number of famous supernova remnants have been detected so far in the slew, 
including, Vela, Puppis-A, N132D and W44. The nature of the slew means
that large areas of the remnants are imaged (Fig.~\ref{fig:snr}) and
detailed two-colour maps of the big remnants will be built up as the 
slew density increases.

\end{document}